# X-rays from Cepheus A East and West

Steven H. Pravdo

Jet Propulsion Laboratory, California Institute of Technology

306-431, 4800 Oak Grove Drive, Pasadena, CA 91109; spravdo@jpl.nasa.gov

and

Yohko Tsuboi

Department of Physics, Faculty of Science and Engineering, Chuo University, Kasuga 1-13-27, Bunkyo-ku, Tokyo 112-8551, Japan; tsuboi@phys.chuo-u.ac.jp

**ABSTRACT**

We report the discovery of X-rays from both components of Cepheus A, East and West, with the *XMM-Newton* Observatory. HH 168 joins the ranks of other energetic Herbig Haro objects that are sources of $T \geq 10^6$ K X-ray emission. The HH 168 effective temperature is $T = 5.8$ (+3.5,-2.3) x $10^6$ K and its unabsorbed luminosity is 1.1 x $10^{29}$ erg s$^{-1}$, making it hotter and less luminous than other representatives of its class. We also detect prominent X-ray emission from the complex of compact radio sources believed to be the power sources for Cep A. We call this source HWX and it is distinguished by its hard X-ray spectrum, $T = 1.2$ (+1.2,-0.5) x $10^8$ K, and complex spatial distribution. It may arise from one or more protostars associated with the radio complex, the outflows, or a combination of the two. We detect 102 X-rays sources; many presumed to be pre-main sequence stars based upon the reddening of their optical/IR counterparts.

# 1. INTRODUCTION

Cep A is star formation region (Sargent 1977) at a distance of ~730 pc (Johnson 1957). It consists of two main H II regions, Cep A East and Cep A West (Hughes & Wouterloot 1982). The eastern region is resolved into several compact radio sources (Hughes & Wouterloot 1984), radio jets (Rodríguez et al. 1994), and masers (Blitz & Lada 1979). The western region also contains compact radio sources (Hughes & Moriarty-Shieven 1990) and Herbig Haro (HH) 168 (GGD 37, Gyulbudaghian et. al. 1978). The latter shows the characteristics of other energetic HH optical condensations with high excitation and linewidths (Hartigan, et al. 1986) and high velocities and proper motions (Lenzen 1988). Hartigan & Lada (1985) identified several optical knots within HH 168 that were resolved and further analyzed with Hubble Space Telescope (HST) spectroscopy (Hartigan, Morse, & Bally 2000).

An enduring question is the identification of the power source or sources in this active region. Much of the attention has focused on two compact radio sources: HW2 and HW3d (Rodríguez et al. 1980, Hughes & Wouterloot 1984). Torrelles et al. (1985) estimate that each source harbors a ZAMS B1 star to account for the ~2 x $10^4$ $L\odot$ in infrared luminosity (Beichman, Becklin, and Wynn-Williams 1979). Polarization results point toward HW2 as the power source (Casement & McLean 1996). However, whether these objects power HH 168 (Lenzen, Hodapp, & Soff 1984) is a matter of contention. Lenzen (1988) measured the proper motions for 6 of the HH 168 knots and concluded that the vectors are directed away from HW2 and 3, "localizing the center of acceleration near this position." Alternatively, Garay et al. (1996) suggest that an ultracompact, variable radio source within the HH 168 HW knot powers the other knots in their east-to-west motion. In this paper we weigh in with the X-ray evidence.

# 2. OBSERVATIONS AND ANALYSIS

## 2.1 X-ray Sources

We observed Cep A with the EPIC cameras of the *XMM-Newton* Observatory on 23 August 2003 for 43.9 ksec. The cameras were in the standard imaging mode. We analyzed the data using the *XMM-Newton* Science Analysis System (SAS, http://xmm.vilspa.esa.es/external/xmm_user_support/documentation/sas_usg/USG/node1.html) and other standard software packages discussed below. EPIC is comprised of three cameras, PN, MOS1, and MOS2. We filtered the raw events data with canonical energy, event-type, and high count-rate criteria. This resulted in net observing times of 37.5 ksec for PN and 43.1 ksec for MOS1 and MOS2. The results of all cameras are consistent. We display spectral data using PN because it has a higher count rate, and spatial data using MOS because their smaller pixel size, 1.1" vs. 4.1", allows better sampling.

We performed source detection on all the data in two energy bands: 0.2-1 keV ("soft") and 1-10 keV ("hard"). A mosaic image of the filtered data from all three cameras is shown in Figure 1 with the detected sources superimposed. The 24 soft sources are shown with red boxes and the 85 hard sources are shown with blue circles. The spot size diameter for a point source viewed with the *XMM-Newton* telescope and camera is about 5". We therefore use this diameter for the source symbols in Figure 1. Seven of the sources appear in both bands, where the criterion for a match between the bands is a positional offset of <2.5". The average and standard deviations of the offsets between the hard and soft source matches is RA = -0.2±0.3" and Decl.= -0.2±1.0". Thus the total number of distinct X-ray sources detected is 102.

We compared the 102 X-ray sources with the USNO B1 (Monet et al. 2003) and 2MASS catalogs. The statistical results are shown in Table 1, where the criterion for a match between the X-ray source and the catalog source is again a positional offset of <2.5". Table 2 lists the sources, their X-ray properties, and their optical and infrared magnitudes. Figure 2 shows the *JHK* diagram for the 62 sources detected. The solid curve in the lower left is the main sequence while the dashed curves show the lines of reddening. The majority of the objects fall along these reddening curves in the locations of pre-main-sequence stars (PMS) including weak-line T Tauri stars (WTTS) and classical T Tauri stars (CTTS). XMMPT J225454.9+615936 (see below) falls in the main-sequence region. A number of outlying sources are labeled and discussed below.

Thirty-nine sources have neither optical nor IR counterparts. Most of these, 29, are detected in the hard band only, and 10 are detected in the soft band only.

### *2.2 HH 168*

The source detection algorithm found two soft X-ray sources within HH 168. Figure 3 shows the softband image in the left panel. Note that the sources are predominantly soft X-rays sources as they are not apparent in the harder X-ray image (right panel). The two sources, XMMPT J225607.1+620159 and J225604.7+620207 are labeled X1 and X2, respectively. Figure 4 shows an Hα image of the region taken with Hubble Space Telescope's WFPC2. We have superimposed on the image the X-ray contours and the locations of the soft X-ray sources (boxes). The prominent HH knots, S, D, E, and HW (Hartigan & Lada 1985) are also labeled. The X-ray emission is highest at the southeastern source, XMMPT J225607.1+620159, which is detected with a statistical significance of 4 σ in each of the three cameras, and a net of 91 counts. The northwestern source, XMMPT J225604.7+620207, is detected at 4, 2, and 3 σ in the PN, MOS1, and MOS2 cameras, respectively, and a net of 58 counts. There is no evidence for intensity variability in these sources.

The peaks of the X-rays emission within HH 168 are significantly offset by several arcseconds from the brightest Hα knots. They are located on the edges of knots E and D. Another smaller enhancement is located near knot HW. The X-ray contours in Figure 4 suggest that these two sources are part of a larger extended emission that covers the entire region of enhanced Hα emission. We compared the radial profiles at the same energies of the soft X-ray emission with that of an expected neutron star point, RXJ171012.3-28075 (Maeda 2004), and find the data are consistent with extended emission (Figure 5, left panel). We treated the emission as extended and extracted events from a circular region (18" radius) containing both sources to perform spectroscopy with the XSPEC program. There were 150 net counts using either of two chosen background regions, east and west of the sources. We fit the spectrum with models containing interstellar absorption ("WABS") and continuum emission-- either thin thermal bremsstrahlung ("MEKA-L") or a powerlaw. The thermal model with a temperature of T = 5.8 (+3.5,-2.3) x $10^6$ K and interstellar absorption of $N_H$ = (4 ± 4) x $10^{21}$ H cm$^{-2}$ yields the best fit (see Table 3).

## 2.3 HWX

A hard-spectrum and non-point-like X-ray source, XMMPT J225618.4+620147 (HWX, hereafter), is located near the center of the radio activity in Cep A East. Figure 3 (right panel) shows the smoothed MOS X-ray image in the 1-10 keV band and the detected sources. Note that the reverse of the HH 168 situation holds: HWX is a predominantly hard X-ray source, undetectable in the softer X-ray image. Figure 5 (right panel) shows a comparison of the HWX linear spatial profile along the line shown in Figure 3 with that of RXJ171012.3-28075, a putative point source. Again, the HWX profile is extended and/or contains multiple sources. Figure 6 shows XMM-Newton MOS contours superimposed on the 2MASS $K_s$ image containing HWX. We also show the compact radio sources (small circles). The HWX emission appears to run in a ridge, approximately southeast to northwest that encompasses some of the radio sources. HW 3d and HW 9 are nearest to the center of the X-ray emission with separations of ~1.8".

The HWX X-ray spectrum is unusually hard, in contrast with that of the HH 168 (Figure 7). It fits a thermal spectrum with T = 1.2 (+1.2,-0.5) x $10^8$ K with $N_H$ = 8 (+3,-2) x $10^{22}$ H cm$^{-2}$, which corresponds to 36 (+14,-9) visual magnitudes of absorption. The spectrum fits equally well to a powerlaw continuum with α = 0.3. The X-ray light curve of HWX was measured within a radius of 12" in the 1-12 keV band. There is an indication of variability at the 95% confidence level.

## 2.4 Other Sources

XMMPT J225454.9+615936 is a source that was probably detected with the *ROSAT* HRI (Zombeck et al. 1990) on 17 December 1996. Since the HRI observation was 15' off-axis, its PSF of several arcseconds allows for the identification of this isolated X-ray source despite its 3.6" separation from the *XMM-Newton* position. The colors resemble an M0V star (see Table 3). If it is an M0V, its distance is ~130 pc, making it a foreground object. Its *XMM-Newton* spectrum is relatively soft with T = 7.0 (+1.2,-4.6) x $10^6$ K. The 0.1-10 keV X-ray intensity was ~1 x $10^{-13}$ ergs cm$^{-2}$ s$^{-1}$ (0.1-10 keV) from HRI, and ~2 x $10^{-14}$ ergs cm$^{-2}$ s$^{-1}$ from *XMM-Newton*, evidence of variability by a factor ~5. There is no evidence for variability or periodicity, however, in the current data alone. If XMMPT J225454.9+615936 is at 130 pc, its X-ray luminosity is 0.4-2 x $10^{29}$ ergs s$^{-1}$, on the high end of the luminosity function for this spectral type (Hünsch et al. 1999).

The two brightest *XMM-Newton* sources are associated with the optical stars HL 8 = J225601.6+620316 and HL 9 = J225604.3+620303. Hartigan & Lada (1985) found these stars to be weak Hα emitters and measured R, I = 15.75, 13.98 and 17.51, 15.49 for HL 8 and 9, respectively, 0.5-1.5 magnitudes fainter than the USNO B1 values (Table 2). Based upon their $M_J$ and *J-K* values these may be K-M stars. The single-component X-ray spectrum of HL 8 fits a power law model far better than a thermal model (reduced $X^2$ of 1.6 compared with 3.8). Yet, for a stellar spectrum there is no good physical interpretation for a power law spectrum with photon index of ~2.3. There is no apparent variability of periodicity in the X-rays that might support a non-thermal origin. A two-component thermal model however also fits. Such a model has in the past been used to interpret PMS spectra (e.g. Feigelson et al.2002). HL 9 fits a thermal spectrum and also appears to be a PMS star.

The *XMM-Newton* source with the brightest optical/NIR counterpart is XMMPT J225643.4+620730. It may also be a foreground object with a soft X-ray spectrum (Table 3) although its colors are reddened. Its X-ray luminosity is ~ 6 x $10^{29}$ erg s$^{-1}$ at the distance of Cep A.

The two reddest sources are labeled in the *JHK* diagram (Figure 2). XMMPT J225620.4+620221 is a soft X-ray source, identified as HL 28 (Hartigan & Lada 1985), an unremarkably red star in the optical. XMMPT J225613.2+620059 is a hard X-ray and IR source with no optical counterpart.

# 3. DISCUSSION

### 3.1 X-rays from HH 168

We conclude that HH 168 joins the growing list of HH objects that are soft X-ray sources, despite that challenging morphology discussed below. The HH objects that have been detected--HH 2H  (Pravdo et al. 2001), HH 154 (Favata et al. 2002), and HH 80/81 (Pravdo, Tsuboi, & Maeda 2004) are the ones with highest-velocities, highest optical excitations, and are associated with compact 6-cm radio emission sources. HH 168 qualifies with respect to velocity (Lenzen, Hodapp, & Solf 1984), excitation (Hartigan et al. 1986, Hartigan, Morse, & Bally 2000), and radio emission (Hughes & Moriarty_Schieven 1990).

The HH 168 X-ray emission appears to be extended, although the spatial resolution of *XMM-Newton* is not sufficient to resolve weak point sources 1-2" apart. The X-ray contours in Figure 4 show a good correlation between the extent of the X-ray and Hα emission. However, whereas in HH 80/81 there is a clear positional correlation between peaks of X-ray and Hα intensity (Pravdo, Tsuboi, & Maeda 2004), the X-ray peaks in HH 168 are offset from those in Hα by ~8''. Knot proper motions of ~0.03"/yr (Lenzen 1988) can not account for this since the Hα observations were made in 1998. Furthermore, HH 168 X-rays do not appear at the leading edges of the westerly moving knots, in contrast with the HH 2H emission (Pravdo et al 2001). On the other hand there is no reason to associate the X-rays with another source, for example, the star HL 14, about 5" north of the stronger X-ray peak.

The luminosity of the composite HH 168 X-ray emission is 1.1 x $10^{29}$ erg s$^{-1}$ assuming the observed low-energy absorption is interstellar. This is the lowest of all the observed HH objects. If the X-rays arise in a shock with the measured X-ray temperature, then the inferred shock velocity is $v_s$ ~ (T/15)$^{1/2}$ = 620 (+170,-140) km s$^{-1}$, larger than the optical knot linewidths and only marginally consistent with the ~475 km s$^{-1}$ linewidth measured by Hartigan et al. (1986) in the HW knot to the southeast edge of the X-ray emission contour. The temperature is higher than that measured in the other HH objects. Closest is the temperature of HH 154, T ~ 4 x $10^6$K (Favata et al. 2002), modeled as originating in the inner portions of the outflow (Bally, Feigelson, & Reipurth 2003).

Figure 8 shows a plot of X-ray vs. radio emission for the known X-ray HH objects. HH 168 *in toto* appears to be underluminous in X-rays relative to its radio emission (Garay et al. 1996). We also show the function $L_X$ ~ $L_R^{1.4}$, the best fit for these data. The fit has a spectral index error of ±0.5 and so is marginally consistent with a linear relationship or an index of 1.24 that Güdel (2002) shows is appropriate for $L_X$ vs. $L_R$ of stars, wherein the energy arises from the magnetic fields.

Applying the model of Raga, Noriega-Crespo, & Velázquez (2002) to HH 168, we find that the X-ray emission regions are nonradiative with a pre-shock number density, $n_0$ ~ 10 cm$^{-3}$, where we have divided the emission among ~2 hot spots each with the volume of the HH 168 knot E.

Why are the HH 168 X-rays both underluminous and offset from the bright optical and radio regions? The answer may lie in the complex geometry of the region and the tangled locations of possible power sources. Ho, Moran, and Rodríguez (1982) conclude that ambient density anisotropies in Cep A lead to outflows that are not well collimated. One can imagine holey Swiss-cheese-like regions from whence the X-rays penetrate in varying amounts, leading to an extended, spatially varying flux profile. Since different density regions preferentially emit at radio, optical, and X-ray wavelengths, the randomization of the emitting and absorbing masses along the lines of sight could create the observed lack of correlation.

### 3.2 The Nature of HWX

Radio observations of Cep A East show many compact radio sources; some with quasi-linear structure (Hughes & Wouterloot 1984) and some that are variable (e.g., Hughes 1991). Hughes (1988) identified these as PMS stars based upon their variability and association with masers. HWX is located at the radio "hub" identified by many observers as the location of the power source or sources of, at least, Cep A East. Figure 6 shows there is an anticorrelation between the HWX X-rays (and most of the radio sources) and the nebular emission, suggesting that X-rays to the northeast are absorbed, *i.e.*, in the direction of HW 4-6. The hard, absorbed HWX spectrum (Figure 6), is in keeping with the idea that X-rays emerge from the edge of a highly absorbed region. The hard spectrum also distinguishes HWX despite its possible extent, from HH 168 and other HH sources that are believed to originate from diffuse jets. The "blankfield" X-ray source, XMMPT J225622.0+620204 emerges from the IR nebula at the end of the string of radio sources extending to the northeast, ~7" from IRS 6d (Goetz et al. 1998).

The peak of the complex HWX emission (right panels of Figures 3 and 5) is consistent with a single point source with the remaining emission due to either additional point sources and/or diffuse emission. Both HW 3d and HW 9 lie within our 2.5" criterion for an association. Torrelles et al. (1998) resolved HW 3d into four radio continuum sources. Of these, (i), (ii), and (iv) are coincident with HWX. HW 3d(ii) is a dominant source associated with $H_2O$ and OH masers (Migenes et al. 1992) and is believed by Torrelles et al. (1998) to be the second youngest in the Cep A region. The HW 3d(ii) coincidence with masers and its positive spectral index (Garay et al. 1996) suggest that it contains a young stellar object (YSO) with our detected hard X-ray emission. HW 9 is variable, appearing and disappearing over ~100 days (Hughes 1991). Hughes (1991) concluded that its radio emission was consistent with gyrosynchrotron emission from a region of ~1 AU, temperature ~$10^8$K, and magnetic field of ~200 G. He interpreted the high electron temperature as a result of magnetic reconnection. However, we note that the thermal emission from such a region is only a few percent of the HWX emission. Other nearby radio sources, including HW 2, also lie in an area of X-ray emission offset from the main peak (Figure 6). The reason that the highly energetic HW 2, often fingered as the Cep A power source, is not more prominent in X-rays may be because it is more highly absorbed than even HWX. Torrelles et al. (1993) estimate a molecular hydrogen column density of $\sim 10^{24}$ cm$^{-2}$, or $N_H \sim 2$ x $10^{24}$ cm$^{-2}$, $\sim 20$ times higher than the HWX measurement.

The HWX X-ray spectrum is highly absorbed with an $N_H$ corresponding to $A_V \sim 36$ mag (Gorenstein 1975), typical of the high absorptions invoked to hide the stellar (e.g. Lenzen, Hodapp, & Solf 1984) or protostellar sources (e.g., Imanishi, Koyama, & Tsuboi 2001, Tsuboi, Hamaguchi, & Koyama 2001). The inferred unabsorbed X-ray luminosity, $\sim 1.6$ x $10^{31}$ erg s$^{-1}$, is larger than the typical values for low-mass protostars in quiescent phases, 1 x $10^{29}$- 4 x $10^{30}$ erg s$^{-1}$ for Class I (Imanishi et al. 2001), and $\sim 10^{30}$ erg s$^{-1}$ for Class 0 candidates (Tsuboi et al. 2001), though it is smaller than that of the extreme protostars SSV 63 E+W in quiescence, $\sim 1$ x $10^{32}$ erg s$^{-1}$ in total (Ozawa et al. 1999). Flaring low-mass protostars can reach the HWX level with X-ray luminosities of $10^{30}$-$10^{32}$ ergs s$^{-1}$ (e.g. Imanishi et al. 2001). On the other hand, the luminosity is also on the high-end for high-mass embedded YSOs (e.g. Kohno, Koyama, & Hamaguchi 2001). The extremely high temperature of HWX (100 million K) is only seen in large flares in low-mass stars (e.g. Tsuboi et al. 1998, Imanishi et al. 2001) or the time-varying emission from plasma from the high-mass YSO IRS 2 in Monoceros (Kohno, Koyama, & Hamaguchi 2001). All these are indicators that HWX is in a high state of X-ray activity, whether it is low mass or high mass.

Another noteworthy discovery is a hard X-ray ridge along the chain of HW 7a-d (Figures 3 and 6). The HW 7 chain is interpreted from the radio spectrum as material shocked by a jet. HW 7d, with a peculiar velocity of 300 km s$^{-1}$, appears to be at the head of the jet (Hughes 1993). Hughes (1993) suggests that HW 9 could be the power source, while Garay et al. (1996) choose HW 3d. Again, these are the two sources with the closest association to HWX. This chain is also along the "high velocity (HV)" outflow seen from near HW 2 to HW 7 in infrared line emission (Goetz et al. 1998). This X-ray ridge cannot be seen in the soft band image (Figure 4) indicating that it arises from a hard and absorbed X-ray spectrum like HWX. If the ridge is as hard as HWX, an origin from a high-velocity shock would require jets moving at $\sim 2000$ km s$^{-1}$, much higher than that the observed motion of HW 7d. Hughes (2001) speculates that if HW 7d originated in HW 3d (or its neighbors) then its age is $\sim 340$ years. If HWX is associated with this power source then we may be viewing a Class 0 protostar relatively early in its $\sim 10^4$ y lifetime (Feigelson & Montmerle 1999).

A final morphological point to note is that the terminus of the Goetz et al. "extreme high velocity (EHV)" outflow is in the nebula proper, near IRS 6d and XMMPT J225622.0+620204 (Figure 6). This source has less net counts than HWX and its spectral parameters, while consistent with those of HWX, are not meaningfully constrained. Its unabsorbed luminosity is $\sim 7$ x $10^{30}$ erg s$^{-1}$.

# 4. CONCLUSIONS

We performed the first high-sensitivity and moderate spatial resolution X-ray observation of Cep A and began to untangle the X-ray morphologies of both Cep A East and West. HH 168 is detected in X-rays and has some but not all of the characteristics of previously discovered X-ray HH objects. HWX is located in the region identified at other wavelengths as the location of the Cep A power sources, but the X-ray evidence is ambiguous as to whether we are seeing the sources, the outflows, or perhaps a combination of both. A more secure indication of short-term HWX variability would favor its stellar or proto-stellar origin. If HWX were comprised of several protostars, higher resolution and sensitivity measurements would provide valuable information regarding their X-ray spectra and luminosities. HWX is harder than the HH jets seen in X-rays. However, if HWX were a jet, it may lie closer to its power sources and thus be more energetic with a resulting higher temperature.

**ACKNOWLEDGEMENTS**

The research described in this paper was performed in part by the Jet Propulsion Laboratory, California Institute of Technology, under contract with the National Aeronautics and Space Administration. Y.T. acknowledges partial support from a Grant-in-Aid for Scientific Research of the Ministry of Education, Culture, Sports, Science, and Technology (No. 15740120), and from a Chuo University Grant for Special Research. Y.T. also thanks the Saneyoshi Scholarship Foundation for their financial support. We thank Y. Maeda for useful discussions and for the providing the RXJ171012.3-28075 data prior to publication. We thank S. Snowden, M. Arida, and the HEASARC for extensive help with *XMM/Newton* data analysis. P. Hartigan graciously provided the processed HST image for our use. We also thank the XMM-Newton User Support. This research has made use of the NASA/ IPAC Infrared Science Archive, which is operated by the Jet Propulsion Laboratory, California Institute of Technology, under contract with the National Aeronautics and Space Administration. This research has made use of the SIMBAD database, operated at CDS, Strasbourg, France. Some of the data presented in this paper were obtained from the Multimission Archive at the Space Telescope Science Institute (MAST). The Association of Universities for Research in Astronomy, Inc. operates STScI, under NASA contract NAS5-26555. This publication makes use of data products from the Two Micron All Sky Survey, which is a joint project of the University of Massachusetts and the Infrared Processing and Analysis Center/California Institute of Technology, funded by the National Aeronautics and Space Administration and the National Science Foundation.

**Table 1. Optical and Infrared Matches for Cep A X-ray Sources**

| X-ray Sources | Catalog | Number | RA Offset (") | Decl. Offset (") |
|---|---|---|---|---|
| Soft (0.1-1 keV) | USNO B1 | 12 | -0.3±1.1 | -0.6±0.8 |
| | 2MASS | 12 | -0.3±1.0 | -0.5±-1.0 |
| Hard (1-10 keV) | USNO B1 | 39 | -0.04±1.0 | -0.01±1.0 |
| | 2MASS | 56 | -0.1±0.9 | 0.2±1.0 |

**Table 3: Spectral Parameters of Selected X-ray Sources in the Cep A Field**

| XMM PT Source | Net counts | kT (keV) | $N_H$ ($10^{22}$ cm$^{-2}$) | I (0.2-2 keV) $10^{-15}$ erg cm$^{-2}$ s$^{-1}$ | I (2-10 keV) $10^{-15}$ erg cm$^{-2}$ s$^{-1}$ |
|---|---|---|---|---|---|
| HH 168 composite | 150 | 0.5 (+0.3,-0.2) | 0.4 ± 0.4 | 3.5 | - |
| HWX | 300 | 10 (+10,-4) | 8 (+3,-2) | 0.2 | 131 |
| J225454.9+615936 | 204 | 0.6 (+0.1,-0.4) | 0.5 ± 0.2 | 18 | 1.6 |
| J225643.4+620730 | 160 | 0.4 (+0.3,-0.2) | 0.8 ± 0.4 | 9.2 | 0.36 |
| HL 8 | 934 | 0.3 and 2.2 | 1.8 | 36 | 93 |
| HL 9 | 385 | 4.9 (+10,-2.6) | 0.5 (+0.5,-0.3) | 20 | 68 |

# FIGURE CAPTIONS

1. The composite X-ray image of Cepheus A from three *XMM-Newton* cameras. Red symbols mark the locations of soft X-ray sources (0.2-1.0 keV) while blue symbols mark the locations of hard X-ray sources (1.0-10.0 keV). The box shows the portion of the image enlarged in Figure 3.

2. We show the *JHK* diagram for the infrared counterparts to the X-ray sources in the Cepheus A field. The solid line is the main sequence and the dashed lines are the reddening lines.

3. These are the XMM-Newton MOS maps of the HH 168 and HWX regions. The data were rebinned into 2.4"x 2.4" pixels and smoothed with a 4.8"-sigma Gaussian. On the left is the 0.2-1. keV image. On the right is the 1.-10. keV image. The lines through the source show the axes for which the linear profiles shown in Figure 5 were taken.

4. The 0.1-1.0 keV XMM-Newton X-ray contours (0.47, 0.64, and 0.87 counts per square arcsecond) superimposed on the HST WF/PC image of Cepheus A taken with the Hα filter (Hartigan, Morse, & Bally 2000). The circles are compact radio sources: blue from Hughes & Moriarty-Schieven (1990) and magenta from Garay et al. (1996). The boxes show the locations of the X-ray sources X1 and X2.

5. A comparison of the X-ray spatial radial profiles between, in the left panel: HH 168 in the soft band with the putative point source, RXJ171012.3-28075 (Maeda 2004), and in the right panel: HWX in the hard band and the same point source.

6. The 1-10 keV XMM-Newton MOS X-ray contours (0.24, 0.33, 0.46, and 0.64 counts per square arcsecond) of the Cepheus A East region superimposed on the 2MASS $K_s$ image. We also show the locations of the detected hard X-ray sources (larger circles, 2.5" radius) and compact radio sources (small circles, Hughes & Wouterloot 1984, Hughes 1991, Goetz et al. 1998).

7. A comparison of the soft X-ray spectrum of the composite HH 168 (dashed symbols) with the hard X-ray spectrum of HWX (solid symbols).

8. The relationship between X-ray and radio (6-cm) luminosity for the known X-ray emitting HH objects. The solid line is a linear fit to these data.

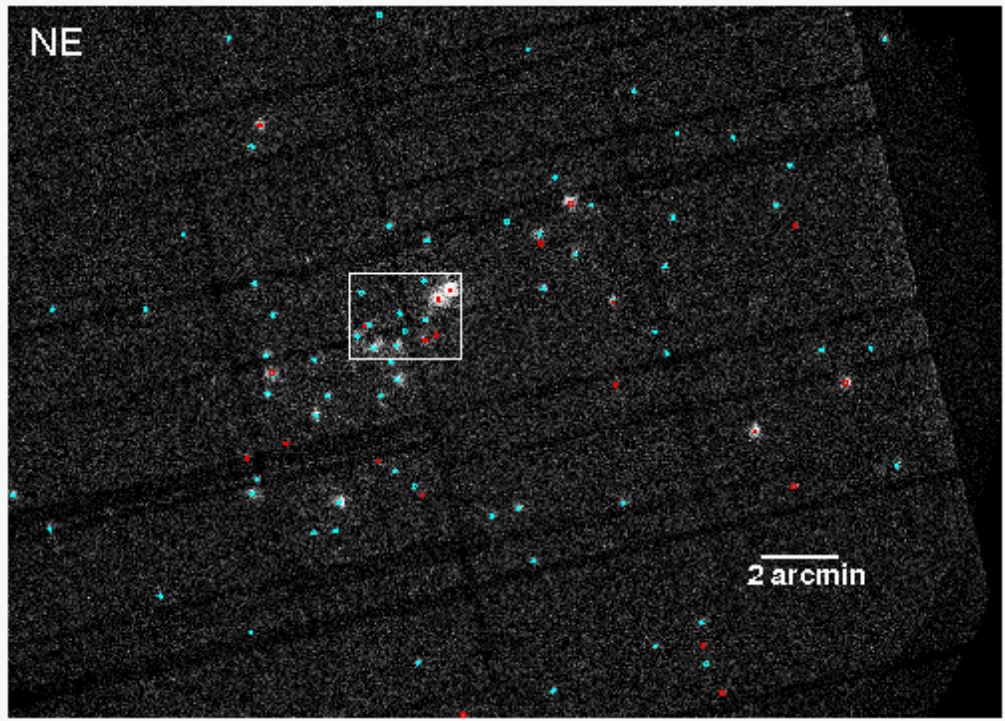

**Figure 1**

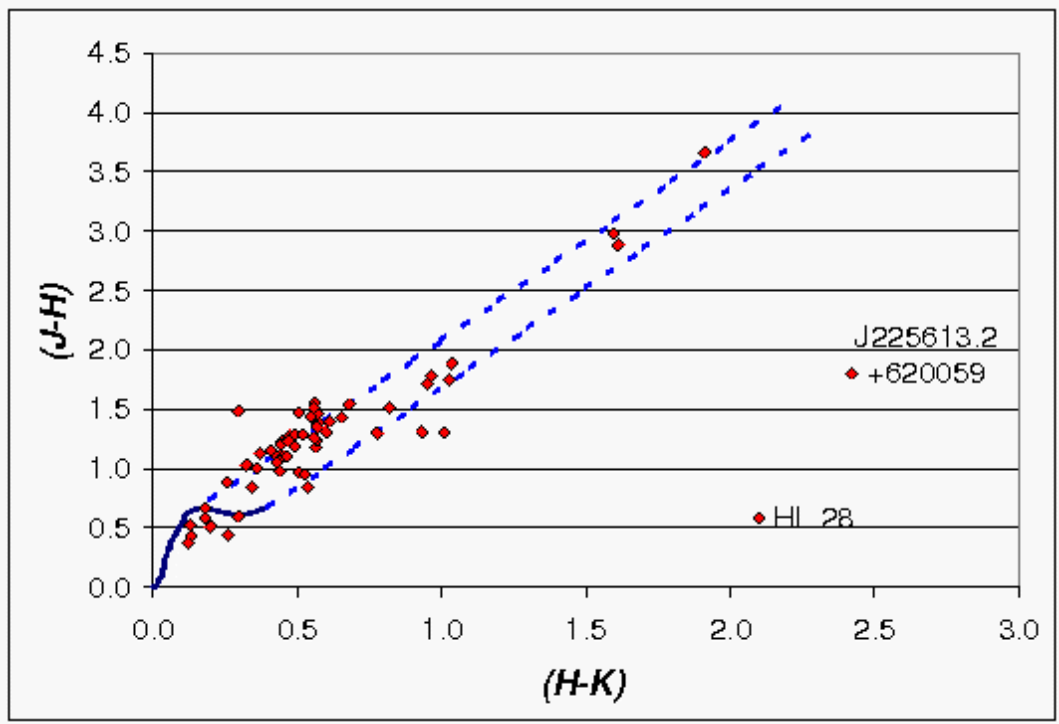

**Figure 2**

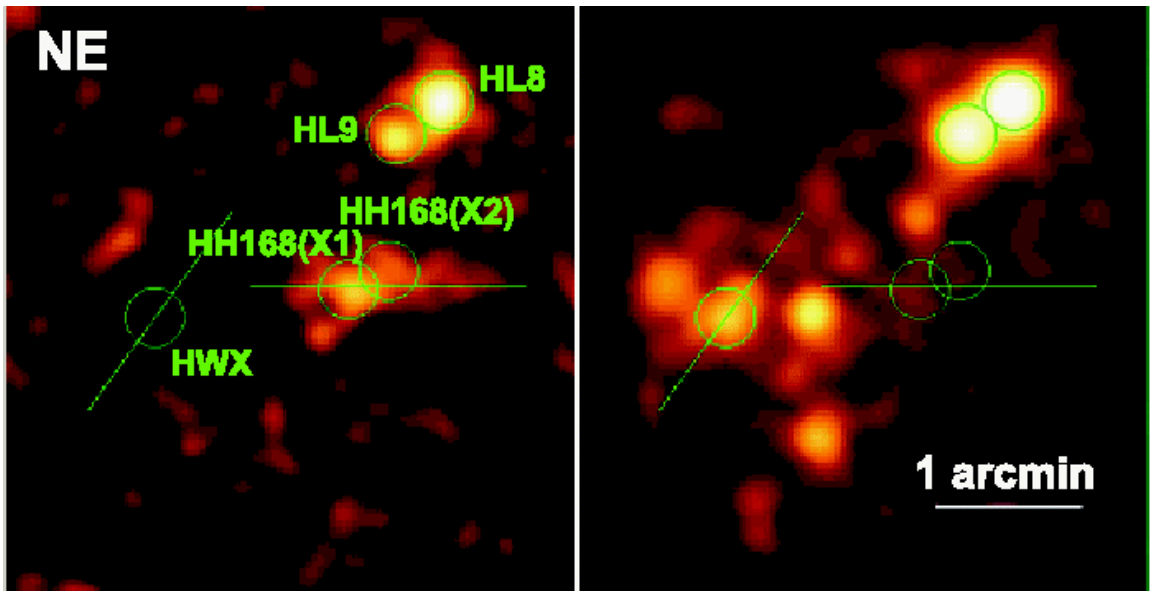

**Figure 3**

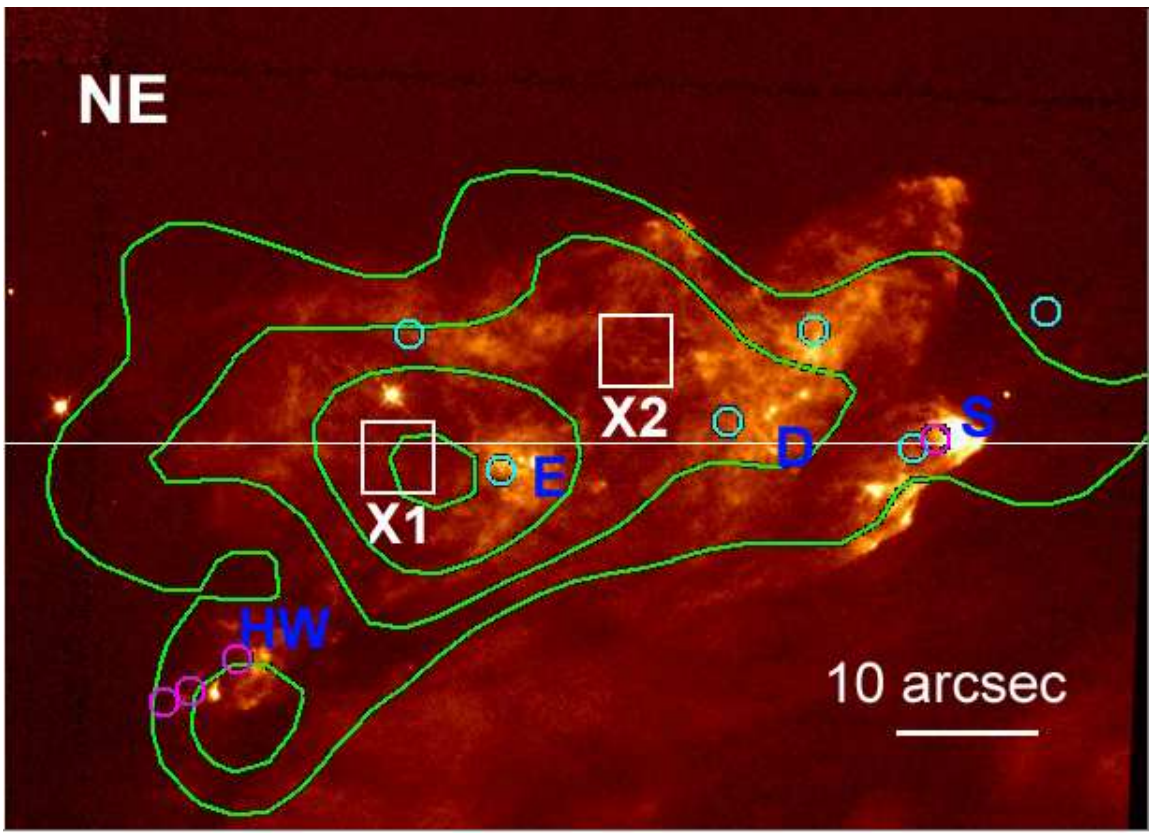

**Figure 4**

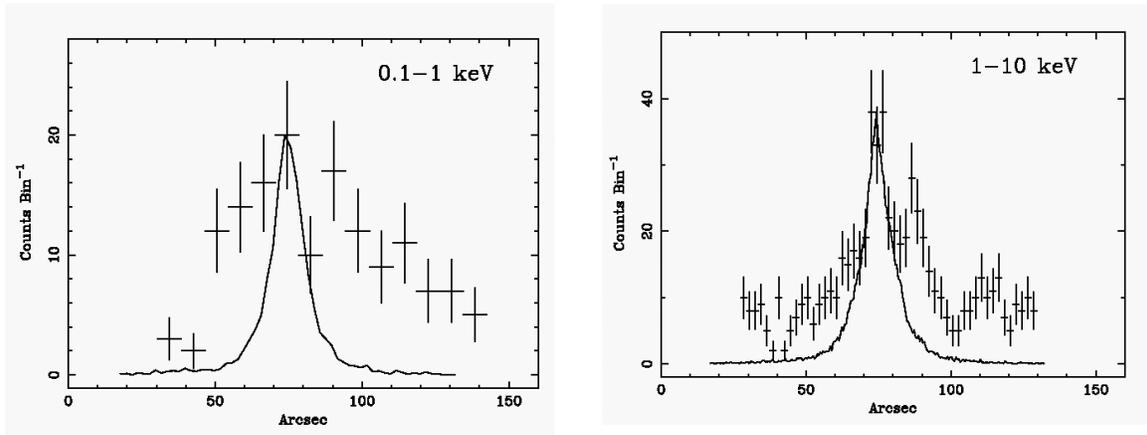

**Figure 5**

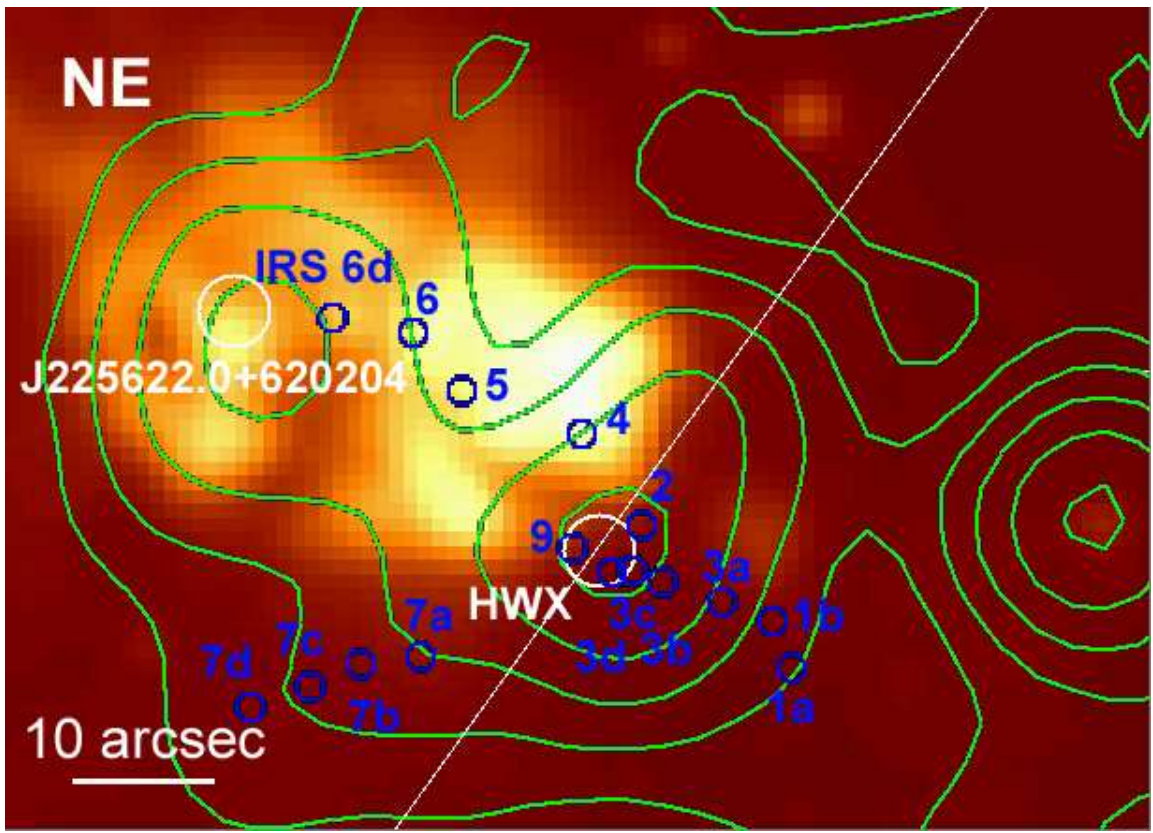

**Figure 6**

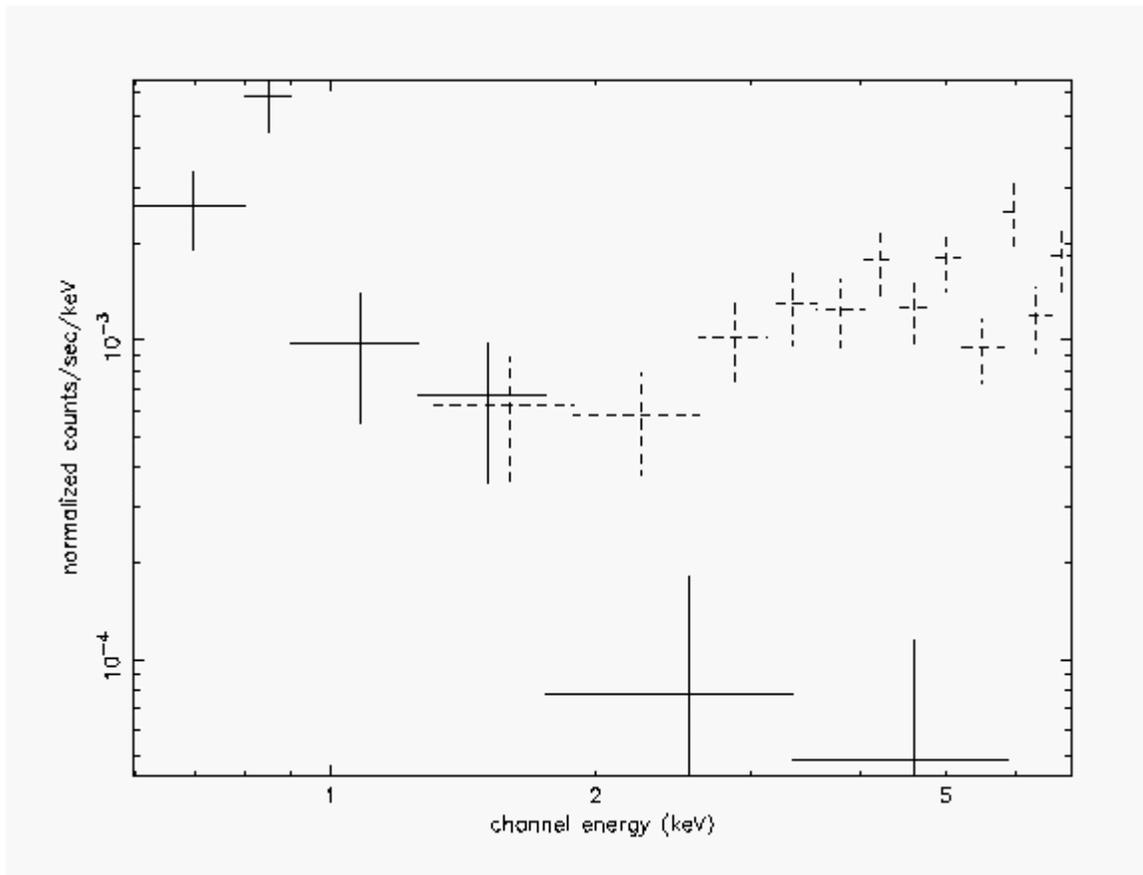

**Figure 7**

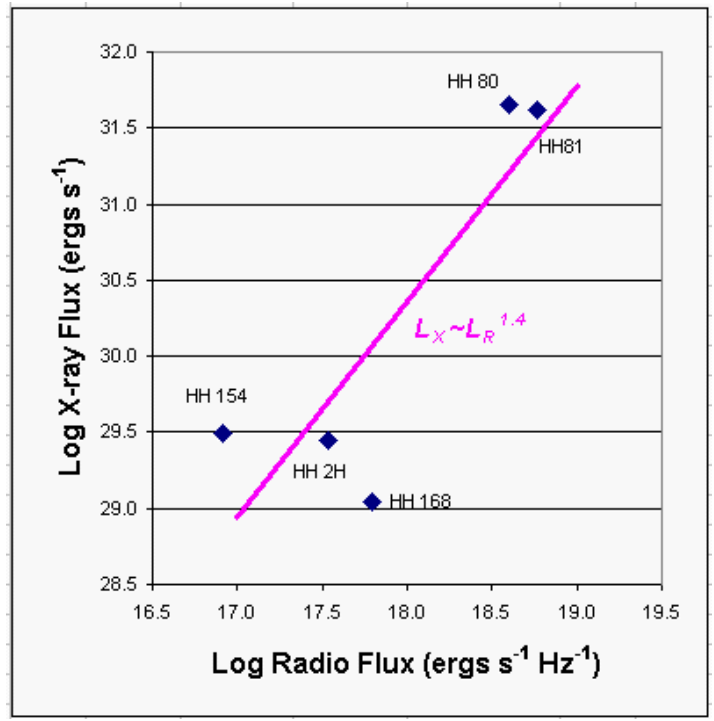

**Figure 8**